\documentstyle[aps,preprint,epsf]{revtex}
\begin{document}
\draft
\preprint{}
\title{Geometric phase for a dimerized disordered continuum:
Topological shot noise}
\author{Prabhakar  Pradhan\cite{pp}}
\address{ Department of Electrical Engineering,
  University of California, Los Angeles, CA 90095}
\author{N. Kumar\cite{nk}}
\address{ Raman Research Institute, Bangalore, 560 080, India}
\maketitle
\date{\today}
\begin{abstract}
Geometric phase shift associated with an electron propagating
through a dimerized-disordered continuum is shown to be 0, or
$\pm \pi$ (modulo 2$\pi$), according as the associated circuit
traversed in the two-dimensional parameter space excludes, or
encircles a certain singularity. This phase-shift is a topological
invariant. Its discontinuous dependence on the electron energy and 
disorder implies a statistical spectral and conductance fluctuation in a 
corresponding mesoscopic system. Inasmuch as the fluctuation derives 
from the discreteness of the phase shift, it may aptly be called a
topological shot-noise.
\end{abstract}

.  \\
.  \\
PACS. 03.65.Bz - Berry's phase.  \\
PACS. 73.23.-b - Mesoscopic systems. \\
PACS. 72.15.-v - Electronic conduction in metals and alloys.  \\
\newpage

The geometric phase, that is the phase shift acquired by a non-
degenerate eigenstate of a quantum-mechanical Hamiltonian as the
latter is cycled adiabatically through a closed circuit in the
space of its parameters, is now a well known and well studied
anholonomy in quantum physics \cite{book}. This phase depends on the
geometry of the circuit traversed, and not on how it is
traversed. Hence, the name geometric phase, as opposed to the
dynamical phase. It was re-discovered and discussed explicitly
in the modern context by Berry \cite{berry},(hence also known
as the Berry phase after him) who had envisaged it for a dynamical
sub-system of interest with {\em fast} degrees of freedom, coupled
to and enslaved by another sub-system of relatively {\em slower}
degrees of freedom (the {\em parameters}), with the total system
wavefunction remaining single-valued, of course.  The idea has
since been generalized to the case of non-adiabatic
evolution \cite{anand}, with the only condition that the evolution be
confined to the same Hilbert space.  The geometric phase in general
varies continuously with the geometry of the parametric circuit
traversed. Thus, e.g., in the case of a spin-1/2 object (an
electron, say) in an external magnetic-field vector cycled over
a cone, the geometric phase acquired by the spin eigenstates is
$\pm$1/2 times the conical solid-angle so subtended \cite{berry}.
There are, however, interesting exceptions where  the geometric phase
can be a topologically invariant quantity having only discrete
allowed values.  A case in point is the Zak \cite{zak} phase acquired by
the Bloch wave for a one-dimensional one-band periodic system where the
parameter, namely the wave vector $\vec q$ is cycled over the closed
reciprocal space (the first Brillouin zone), with $\vec q$ and $\vec q+ \vec G$
identified. (Here $\vec G \equiv$ shortest reciprocal lattice vector). In
the present work, we report yet another case of a topological phase
shift that appears naturally when we consider an electron
propagating in a dimerized-disordered one-dimensional lattice,
treated here in the continuum limit.
Here, the time-independent Schr\"{o}dinger equation for the
wave-amplitude can be transformed to the \underline{spatial} evolution equation
for a pseudo-spin-1/2 object in a pseudo-magnetic field generating  a
two-dimensional parameter space. The evolution, however, turns
out to be {\em non-unitary}. Following the recently developed
procedure for non-unitary evolution \cite{sham,nmk}, we have
calculated the geometric phase for a circuit and found it to be $0$, or
$\pm\pi$ according as the circuit traversed excludes, or encircles a
certain singular point in the parameter space. This topological phase
depends discontinuously on disorder and on the electron energy,
and its discreteness is, therefore, expected to introduce a
shot-noise characteristic in the statistics of spectral and
conductance fluctuations for certain mesoscopic systems. Observable
and calculable consequences for disordered mesoscopic systems are
discussed. 

Consider the Schr\"{o}dinger equation for the wave amplitude
$\psi_n$ for a 1D tight-binding system given by
\begin{mathletters}
\begin{eqnarray}
  i\hbar \frac{\partial \psi_n}{\partial t} &=&
  V_n^E \psi_n - \nu(\psi_{n+1} +\psi_{n-1}),\\
  i\hbar\frac{\partial \psi_n}{\partial t}  &=& 
  V_n^O \psi_n - \nu (\psi_{n+1} +\psi_{n-1}),
\label{seq}
\end{eqnarray}
\end{mathletters}
where dimerized-disorder has been introduced through the random
site-diagonal potentials $V_n^E$ and $V_n^O$ for the even and
the odd sites respectively. More specifically, we set $V_n^E =
U_n + \eta _n$, $V_n^O = U_n - \eta_n$ and treat $U_n$ and
$\eta_n$ as statistically independent random variables. This
bipartite structure is essential to our work. We now proceed to
the continuum limit, but without losing the two-sublattice
structure. For this, we define \cite{emery} 
\begin{mathletters}
\begin{eqnarray}
\psi_{2n}(t) &=& i^{2n} (2s)^{1/2} \psi_E(x,t), \\
\psi_{2m+1}(t)&=& i^{2m+1} (2s)^{1/2} \psi_O(y,t),
\label{preserve}
\end{eqnarray}
\end{mathletters}
and let the lattice spacing $s\rightarrow 0$, total site number
$N\rightarrow \infty$, keeping the total length $L = Ns$ fixed
and setting  $2ns = x$, $(2m+1)s = y$. In order to have a
well defined continuum limit, the coupling $\nu$ must also be
scaled as $\nu \rightarrow \infty$, keeping $2s\nu/\hbar =v_F$.
Then, defining $\psi_L = (\psi_E +\psi_O)/\sqrt{2}$ and
$\psi_R = (\psi_E - \psi_O)/\sqrt{2}$, we get
\begin{equation}
i\hbar \frac{\partial}{\partial t}
\left( \begin{array}{c} 
\psi_L  \\  \psi_R 
\end{array} \right) =
\left( \begin{array}{cc} 
        U(x) - i\hbar v_F \frac{\partial}{\partial x}  &  \eta (x) \\
        \eta (x)  &  U(x) + i\hbar v_F \frac{\partial}{\partial x}
\end{array}\right) 
\left( \begin{array}{c}  \psi_L  \\  \psi_R
\end{array} \right). 
\label{tevo}
\end{equation}
 It is clear that the above equation describes the left-going and the
right-going wave amplitudes (also called the left- and
right-movers in the context of a two-component Dirac equation
for a massless particle in 1-space + 1-time dimension) with the
back-scattering admixing them.  Thus, for $U(x)=0=\eta(x)$, the
electron energy $E=\pm\hbar v_Fk$, where $k$ is the wavevector.  We
can rewrite it by introducing a two-component wave amplitude $\psi$ as 
\begin{equation}
 \psi=\left( \begin{array}{c}  \psi_L  \\  \psi_R  
\end{array} \right),
\label{vpsi}
\end{equation}
and the evolution equation Eq.\ref{tevo} then becomes
\begin{equation}
i\hbar v_F {\frac{\partial\psi}{\partial x}} = -(E -
U(x))\sigma_z\psi + i\eta(x)\sigma_y\psi \equiv H_{eff}\psi,
\label{xevo}
\end{equation}
where we have set $i\hbar \frac{\partial}{\partial t} \equiv
E $, the electron energy. This is our main equation and can be
studied for possible geometric phases. In this form the
connection with a pseudo-spin 1/2 evolving spatially in a
pseudo-field is clear. However, the evolution is 
now in {\em space} and is
{\em non-unitary} because of the imaginary "$i$" occurring on the
right-hand side of the equation.

We now proceed to calculate the geometric phase  associated with
the spatial evolution given by Eq.\ref{xevo}. First note that the 
instantaneous eigenvalues of the evolution (non-Hermitian)
Hamiltonian $H_{eff}$ in Eq.\ref{xevo} are:
\begin{equation}
\lambda_\pm = \pm \sqrt{(E - U(x))^2 - \eta(x)^2}
\end{equation}
and the corresponding right-eigenvectors (unnormalized) are:
\begin{equation}
|\pm> =
\left( \begin{array}{c} \frac{ (E-U(x)) \mp  \sqrt{(E-U(x))^2
-\eta(x)^2 }}{\eta(x)} \\ 
1
\end{array} \right), \\
\label{eigenfn}
\end{equation}
noindent with bi-orthogonality between the left- and the
right-eigenvectors. 

Consider now the 1D disordered dimerized system as terminated
asymptotically into two ordered leads that may be joined. We now
consider the evolution of the two-component wave amplitude
$\psi(x)$ as $x$ varies from one ordered lead to the other
through a disordered segment. This will subtend a closed circuit
in the two-dimensional parameter space $R(U(x),\eta(x))$. The
question we are addressing now is {\em whether or not there is a
geometric phase associated with the circuit and its magnitude if
there is any}. Such an anholonomy is expected from the form of
the evolution equation Eq.\ref{xevo} even though the evolution is
non-unitary. The essential feature is the non-commutativity of
the diagonal and the off-diagonal parts of the Hamiltonian
$H_{eff}$.  {\em It is important to note here, that the
circuit in the parameter space depends explicitly on the energy
}$E$. We take the random potentials to vary slowly in the
$x$-space so as to make the evolution adiabatic in the sense of
the geometric phase.

\begin{figure}
\epsfxsize=12cm
\centerline{\epsfbox{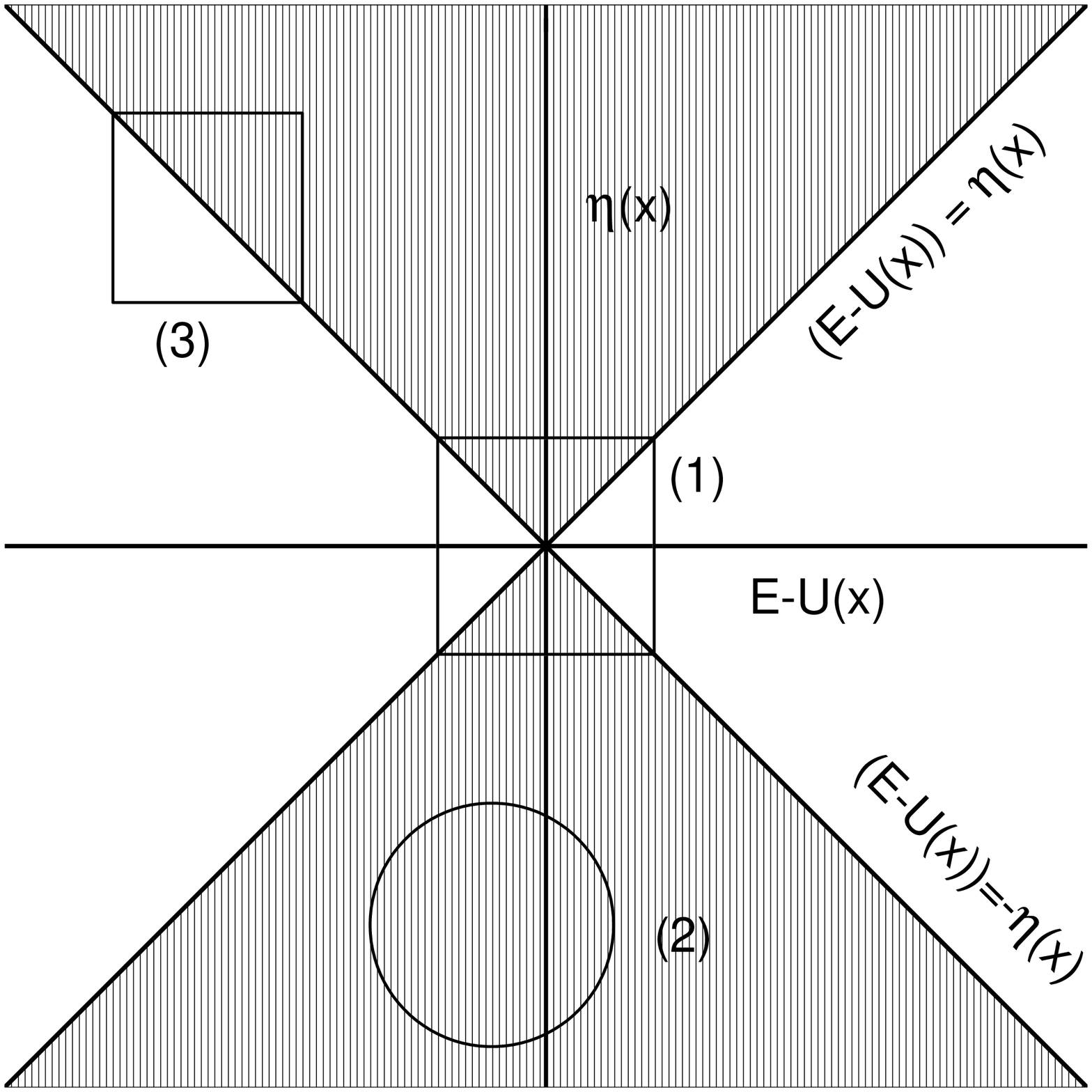}}
\vspace{.8cm}
\caption{Parameter space $R((E-U(x),\eta(x))$ for the
 adiabatic circuit.
 Shaded portion marks the regime where geometric phase shift exists, that
is, $|(E-U(x))| < |\eta(x)|$.}
\label{pspace}
\end{figure}

As noted earlier, a first principle calculation of the geometric
phase is given in Ref.\cite{berry} for an adiabatic and unitary
evolution of the Hamiltonian on a closed circuit in a parameter
space. Calculation of the geometric phase was later generalized
for the non-unitary evolution \cite{sham,nmk}.
The geometric phase  $\gamma$  for an adiabatic evolution in terms of the
instantaneous right-eigenstate is given by
\begin{equation}
\gamma_{\pm } = i\oint \frac{<\pm |\partial\vec R \pm >}
{<\pm |\pm >} d\vec R \,, 
\label{phsdef}
\end{equation}
where $\vec R$ traces the circuit in
the two-dimensional parameter spaces $R((E - U), \eta )$.
Now, from Eq.\ref{phsdef}, the values of the geometric phases for the
instantaneous eigenvectors $\mid+>$ and $\mid->$ are, 
\begin{eqnarray}
\gamma_{\pm} &=& \mp \oint \frac{1}{2}\left[ \frac{1/\eta
\,\,d(E-U)}{\sqrt 
{1-(E-U)^2/\eta^2}} \right. \nonumber \\
&&\left. - \frac{(E-U)/\eta^2 \,\,d\eta}{\sqrt
{1-(E-U)^2/\eta^2}} \right].  
\label{phsexp}
\end{eqnarray}
From Eq.\ref{phsdef}, it is clear that in order to have a non-zero
geometric phase, the "instantaneous" eigenvector has to be
complex. The eigenvector expressions in Eq.\ref{eigenfn} give
$\mid\pm >$ to be complex if $\mid\eta(x)\mid > \mid E-U(x)
\mid$. In Fig.\ref{pspace} we have drawn the parameter space
$R((E - U(x)),\eta(x))$ with typical circuits
marked (1), (2) or (3) . The region
$\mid\eta(x)\mid > \mid E - U(x)
\mid $ is shown in shade, where eigenvectors are complex and the
geometric phase shift arises. In the non-shaded region the
eigenvectors are real and have no geometric phases associated
with them.  Straightforward application of Stokes theorem shows
that the total geometric phase associated with a closed circuit
in the parameter space is zero except for the case of circuits
encircling the singular point $E-U = 0 $, $\eta = 0$.
Also, in the latter case, the phase shift is independent of the
particular circuit  traversed. We have, therefore, chosen a
simple circuit  marked (1) as shown in Fig.\ref{pspace} for computational case.
We get (setting $(E - U)/\eta = z)$:
\begin{equation}
\gamma_\pm = \mp \frac{1}{2}\left[2\sin^{-1} z \mid_{-1}^{+1}\right] =
\pm \pi \,\,\,\,{\rm (mod }\,2\pi).
\label{phase}
\end{equation}
{\em This is our main result.}

    It will be apt at this stage to discuss  the validity of
    Eq.10. In particular, we would like to assure ourselves that the 
  approximations of the continuum limit, of adiabaticity and the
  consideration of localization do not drastically limit the domain
  of validity of the Eq.10. First, the continuum limit (Eq.2) in 
   relation to the singular points  where the right-hand side of the
    Eq.6 vanishes, giving the  topological phase jump.
 In the absence of disorder, the dimer-to-dimer spatial variation
of the two-component wave function for the discrete dimerized lattice
is controlled by the wavevector, $k$, for a given energy eigenvalue, $E$.
With disorder, one has to consider the local wavevector $k(x)$, which vanishes
at the 'turning points' $E-U=0=\eta$, where the local energy eigenvalue
$\lambda_{\pm}$ vanishes. This ensures a slow spatial variation as 
$k(x)s=0$ at these points, where $s$ is the lattice spacing. 
This justifies  our continuum limit at the point $E-U(x)=0=\eta(x)$,
that generates the topological phase-shift. The topological nature
of this phase shift then ensures validity of our treatment in 
the neighbourhood of this singular point. The point to note here is that
the slowness of variation of the wave function, relevant to 
our continuum limit, is over dimer-to-dimer spacing, and {\em not intra-dimer}.
The latter is {\em absorbed} in the two-component nature of our wave function. 
This is quite analogous to the two-component Dirac continuum limit of 
the two-band models, treated in the literature \cite{emery}  
In our case, the dirac equation is massless. Next, we consider the
effect of localization.
 First, it is seen from the Eq.3, that the disorder  $U(x)$ can give only
forward scattering and hence can not be very effective for localization.
It is the disorder $\eta(x)$ that gives back-scattering and is effective
for localization. However, both $U(x)$ and $\eta(x)$ control the path in
the parameter space, and, therefore, whether or not the path encircles 
the singular point. One can thus have weak disorder, so as to have
 the localization length exceed the sample length and still not invalidate
 the continuum limit in view of our discussion  above. Slow variation
 of disorder in space can ensure adiabaticity. Thus, in principle, 
 the parameter space for the validity of our treatment is not constrained
 by conditions for continuum limit, localization and adiabaticity, 
 militating against one another.

 The question now is what observable consequences this geometric
(topological) phase may have. Given that it is the relative,
rather than the absolute phase that has a physical
significance, one has basically two ways of demonstrating the
geometric-phase sensitive effects. Either prepare the quantum
system in a coherent superposition of two instantaneous
eigenstates that pick up  different geometric phases for the
same path traversed in the parameter space (i.e., two states but
one Hamiltonian), or prepare the system in a given initial
(instantaneous) eigenstate and let it evolve along two different
alternative paths traversing the parameter spaces so as to give
an interference between the partial amplitudes (one state and
two Hamiltonians).
The physical significance of the above geometric (topological) phase
is, however, best realized by considering the following situation.
Consider a 1D disordered dimerized conductor in the form of a ring.
Such a system automatically ensures a closed circuit in the parameter
space as discussed above. 
Mathematically this means imposing the periodic boundary
condition. The periodic boundary condition involves the matching of
the phases ---the phase change accumulated around the ring must be a
multiple of $2\pi$. But now this must include the extra
geometric(topological) phase as well. This will alter the eigenvalue
spectrum for the ring and must be included in any spectral reckoning.
In the limit of weak scattering, i.e., disorder parameter $<<$ level
spacing (which can be obtained for small enough ring size, the
mesoscopic system) this extra phase of $\pm \pi$ will cause a level
shift. A rather subtle and experimentally observable effect
associated with this geometric phase is made plausible if we consider
the disorder $U(x)$ and $\eta(x)$ to be weakly and slowly 
modulated in time.
This has the effect of translating the parametric (sample specific)
fluctuation into temporal fluctuations.
This happens naturally in mesoscopic systems at low
temperatures. The modulation depth
can be such that the circuit in the $U(x)$ and $\eta(x)$ parameter
space sweeps across the central singularity ($\eta(x)=0=E-U(x)$)
--- i.e., the geometric phase is cyclically flipped from zero to $\pm
\pi$ {\em discontinuously}. The discontinuous nature of our
topological phase change should lead to a {\em large spectral (level
spacing) and conductance fluctuations manifesting as noise}.
The discreteness of the topological phase shift (as distinct from
the usual continuous geometric phase) should make this the statistical
analogue of a shot-noise.
Finally, we must note that the circuit in the $U(x)$ and $\eta(x)$
parameter space involves energy $E$ explicitly.
Thus, for a given realization of randomness, as $E$ varies we
expect singularities in the energy spectrum at the special values of
$E$ corresponding to the circuit enclosing, or not enclosing, the
center i.e., the singularity of the parameter space.
 This suggests the following experimentally realizable possibility  
 of observing certain phase-sensitive effects  by combining
 our  pseudo-magnetic field with a real Aharonov-Bohm magnetic
 flux, threading the matrial ring connected to two leads. The
 variation of the  energy level  as function of the real flux can
 then  tune it across the singular point causing the phase
 jump. This must manifest as jump in the transmission through the
 ring. Similarly, one should expect jumps in the persistent ring
 currents.    

          We would like to conclude with the following remarks in support of
what is new in our work. Topological phase shift is by itself not a new idea
now. It is a well known particular case of Berry's phase when, e.g. the
magnetic field acting on a  spin is confined to a planar transport
circuit. Indeed, such situations have been realised literally experimentally
involving, e.g. a momentum-dependent magnetic field via the spin-orbit
coupling \cite{geller}. In our work, the pseudo-magnetic field 
arises in a novel way
through the dimerised nature of the medium, and one has to calculate the
topological phase shift for the effectively non-unitary evolution of a
properly defined state vector. Further, the parametric trajectory traced out
by this pseudo-magnetic field, as one goes along the sample length, depends
sensitively on the sample specific realization of the disorder. It is this
sensitivity coupled with the discreteness of the topological phase shift
that generates the shot noise when the sample scans the parameter space
through, e.g. a phononic  modulation. Clearly, other manifestations of this
topological phase are expected, e.g. through its effect on the level spacing
statistics. 
We hope that this work will stimulate experimental work on mesoscopic
rings made from quasi-one-dimensional disordered dimerized materials, 
i.e. material having
bipartite lattice structure. Many conducting pseudo one-dimensional
polymeric systems may belong to this category of materials. 


We thank R. Simon for useful discussions.


\end{document}